\documentclass{article}
\usepackage{graphicx}
\usepackage{amsmath}
\usepackage{natbib}
\usepackage[colorlinks,citecolor=blue,urlcolor=blue,bookmarks=false,hypertexnames=true]{hyperref} 
\usepackage{booktabs}

\usepackage{enumitem}
\setlist[description]{%
  topsep=0pt,               
  font={\bfseries}, 
}

\usepackage[margin=1.2in]{geometry}
\usepackage{lineno}

\begin{document}


\setcounter{secnumdepth}{2}

\title{Timber Volume Estimation Based on Airborne Laser Scanning -- Comparing the Use of National Forest Inventory and Forest Management Inventory Data}

\author{Johannes~Rahlf\thanks{J.~Rahlf (johannes.rahlf@nibio.no) and J.~Breidenbach(johannes.breidenbach@nibio.no)} \and Marius~Hauglin \and Rasmus~Astrup \and Johannes~Breidenbach\footnotemark[\value{footnote}]}

 \date{     Division of Forest and Forest Resources\\
            Norwegian Institute for Bioeconomy Research (NIBIO)\\
            H{\o}gskoleveien 8, 1433 ~{\AA}s,~Norway\\ [2ex]%
    \today 
}

\maketitle

\begin{abstract}

~

\begin{description}
\item [Key message] ~\\Large-scale forest resource maps based on national forest inventory (NFI) data and airborne laser scanning may facilitate synergies between NFIs and forest management inventories (FMIs). A comparison of models used in such a NFI-based map and a FMI indicate that NFI-based maps can directly be used in FMIs to estimate timber volume of mature spruce forests.

\item [Context] ~\\Traditionally, FMIs and NFIs have been separate activities. The increasing availability of detailed NFI-based forest resource maps provides the possibility to eliminate or reduce the need of field sample plot measurements in FMIs if their accuracy is similar.

\item [Aims] ~\\We aim to 1) compare a timber volume model used in a NFI-based map and models used in a FMI, and 2) evaluate utilizing additional local sample plots in the model of the NFI-based map.

\item [Methods] ~\\Accuracies of timber volume estimates using models from an existing NFI-based map and a FMI were compared at plot and stand level. 

\item [Results] ~\\Estimates from the NFI-based map were similar to or more accurate than the FMI. The addition of local plots to the modeling data did not clearly improve the model of the NFI-based map.

\item [Conclusion] ~\\The comparison indicates that NFI-based maps can directly be used in FMIs for timber volume estimation in mature spruce stands, leading to potentially large cost savings.

\end{description}

\end{abstract}

\maketitle
\newpage

\section{Introduction\label{introduction}}

Forest management inventories (FMIs) in the Nordic countries
\citep{Naesset2004d} mainly provide stand-level information in order to
support forest management decisions, while national forest inventories
(NFI) mainly provide statistics for reporting and policy making on
regional to national scale \citep{tomppo2010National, vidal2016national}.
Traditionally, FMIs and NFIs have been completely separate activities but the increasing availability of fine-resolution remotely-sensed 3D-data such as large-scale or even nationwide airborne laser scanning (ALS) campaigns has triggered the creation of detailed national forest resource maps, and as a consequence, the search for synergies between NFIs and FMIs \citep{Kangas2018}.

Fine-resolution 3D data have been used in forest inventories for many years.
Remotely sensed data from ALS or digital aerial photogrammetry allow accurate estimation of forest stand parameters \citep{rahlf2014comparison} to support forest management decisions \citep{kangas2018value}.
In the Nordic countries, ALS is currently the most common method for the acquisition of auxiliary data in FMIs \citep{naesset2014area,  maltamo2020comprehensive}.
Common steps in an ALS-based FMI are
1) manual stand delineation, 
2) stratification of the stands into four or more tree-species and maturity-class specific strata, 
3) ALS data acquisition,
4) measurements of some hundred field sample plots systematically distributed in the strata, 
5) fitting of stratum-specific linking models for timber volume and other response variables, and 
6) estimation of stand-level parameters.
One main outcome of FMIs is a stand map or stand list that includes stand-level information on the dominant species and the estimated timber volume.

NFIs, on the other hand, collect data for regional or national statistics over several years according to a national systematic design.
Such a sampling design results in an overall larger data set of field plot measurements at the large scale but a much smaller sampling fraction than in a FMI.
Nevertheless, earlier studies have explored the use of NFI sample plot data with remote sensing for stand-level estimation of forest parameters.
\citet{McRoberts2008using} achieved promising results using Landsat satellite imagery to bridge strategic inventories and FMIs in Minnesota, USA.
\citet{breidenbach2018unit} compared unit level and area level models and estimators based on NFI and digital aerial photogrammetry data.
\citet{Maltamo2009a} and \citet{tuominen2014nfi} used ALS data and Finnish angle count NFI data with or without additional fixed radius sample plot measurements to estimate stand parameters and assessed estimation accuracy.
While the use of NFI plots produced acceptable estimates and improved accuracies when used together with fixed radius plots, the NFI sample plot designs caused problems for practical application.

Other approaches combining NFI and remotely sensed data focus on the creation of national forest resource maps by wall-to-wall mapping of forest parameters on nationwide scale.
Early examples of such maps linked coarse optical satellite data with NFI sample plots as reference data \citep{tomppo1991satellite,reese2003countrywide, Gjertsen2007}.
In recent years, coverage with fine resolution 3D remotely sensed data has increased drastically.
On one hand, advances in soft- and hardware made it possible to compute large-scale 3D information from aerial imagery by means of digital aerial photogrammetry.
Such 3D data have been used to create forest resource maps covering large regions or nations \citep{breidenbach2016empirical, bohlin2017mapping, rahlf2017digital, waser2017wall, Astrup2019}, using NFI data as reference.
On the other hand, large-scale ALS campaigns have been or are currently being conducted in several countries, manly aiming at the creation of accurate digital terrain models.
While the remotely sensed data used in these maps are often not optimized for forest analyses and varying sensors, acquisition settings and conditions lead to variation in the ALS data \citep{naesset2009effects, Naesset2008, hill2018combining}, the large-area coverage enables the use of a great number of NFI sample plots in the fitting of forest parameter models.
Examples of forest maps based on large-scale ALS data are \citet{nilsson2016nationwide}, \citet{monnet2016wide} and \citet{nord2012estimation}.
In the following, we will denote these data sets \emph{``NFI-based maps''}, and the regression models that link the ALS data with NFI plot data as \emph{``NFI models''}.
NFI-based maps typically contain forest parameters required in FMIs \citep{naesset2014area}.

Studies on the transferability of FMI models among areas
\citep{tompalski2019demonstrating, karjalainen2019transferability}
suggest that models linking ALS metrics with variables of interest are
relatively stable in most cases and often can be transferred between
different areas, but that systematic errors may occur. These results are to be
expected because synthetic estimators, e.g.~estimators aggregating model
predictions, only have small bias, if the local condition corresponds to
the condition in the population used to fit the model  (\citealp[p.~36]{Rao2015}; \citealp{mandallaz2013design}).
Related to transferring models between different areas is the
application of a large-scale model, as used for NFI-based maps,
within a smaller area, where only a few or no sample plots are located.
While error analyses for NFI-based maps have been conducted on stand scale \citep[e.g.~][]{nilsson2016nationwide}, it has not been studied how NFI-based map estimates compare to FMI estimates.
If (synthetic) stand-level estimates from NFI-based maps were similar to estimates from a traditional FMI approach, the NFI-based maps could easily be used in FMIs.
The use of the NFI-based map would then allow considerable cost saving by eliminating or reducing the need for field plot measurements.

The aim of this study is to compare the performance of a model used in a NFI-based map, namely the Norwegian national forest resources map SR16, and models used in a traditional FMI in terms of accuracy of plot-level predictions and stand-level estimates of timber volume of mature spruce stands.
Furthermore, we evaluate utilizing additional local sample plots in the model of the NFI-based map (\emph{``adjusted NFI model''}) in an attempt to improve local model accuracy.

\section{Material and methods}\label{material-and-methods}

\subsection{Study area}\label{study-area}

\begin{figure}
\centering
\includegraphics[height=0.4\textheight]{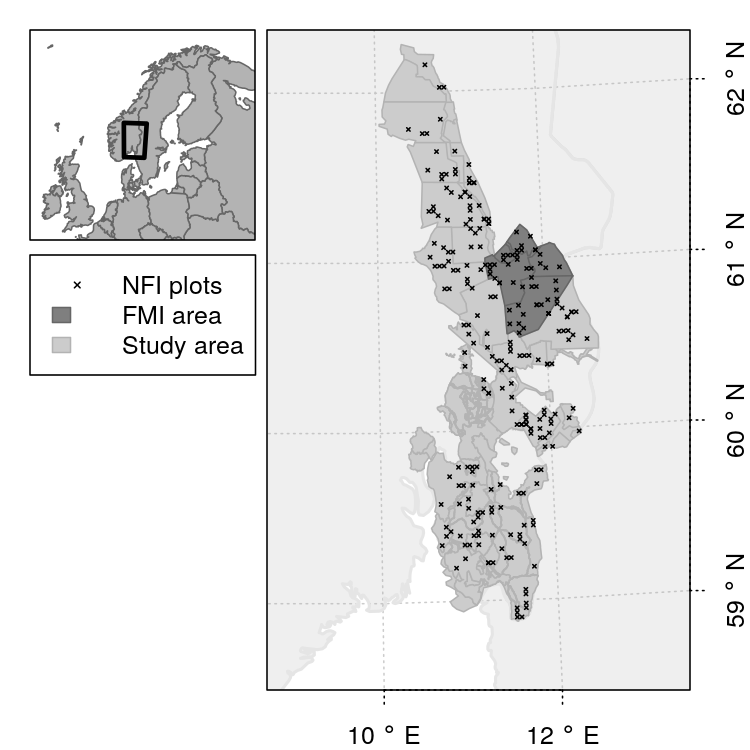}
\caption{Location of the study area and the NFI plots used. Subdivisions within the study area indicate different ALS
projects. The area covered by the FMI is highlighted in dark gray. \label{fig_map1}}
\end{figure}

The study area is in southeastern Norway, covering parts of the counties Innlandet and Viken (Figure~\ref{fig_map1}).
A FMI was conducted within the study area and covers the municipalities Våler and Elverum.
The forests are dominated by Norway spruce (\textit{Picea abies} (L.) Karst.) and Scots pine (\textit{Pinus sylvestris} L.), comprising 52\% and 35\% of the overall timber volume, respectively.
Deciduous tree species are less frequent (13\% of the overall timber volume).
For the current analysis, we focus on the FMI stratum consisting of Norway spruce-dominated stands in the maturity classes ``production forest'' and ``old production forest'', which comprise the oldest and economically most valuable stands.
For simplicity, we will refer to the stratum of interest as \emph{mature spruce forest}.
A description of the Norwegian maturity class system can be found in \citet[Section 5.3.2]{breidenbach2020century}.

\begin{figure}
\centering
\includegraphics[height=0.4\textheight]{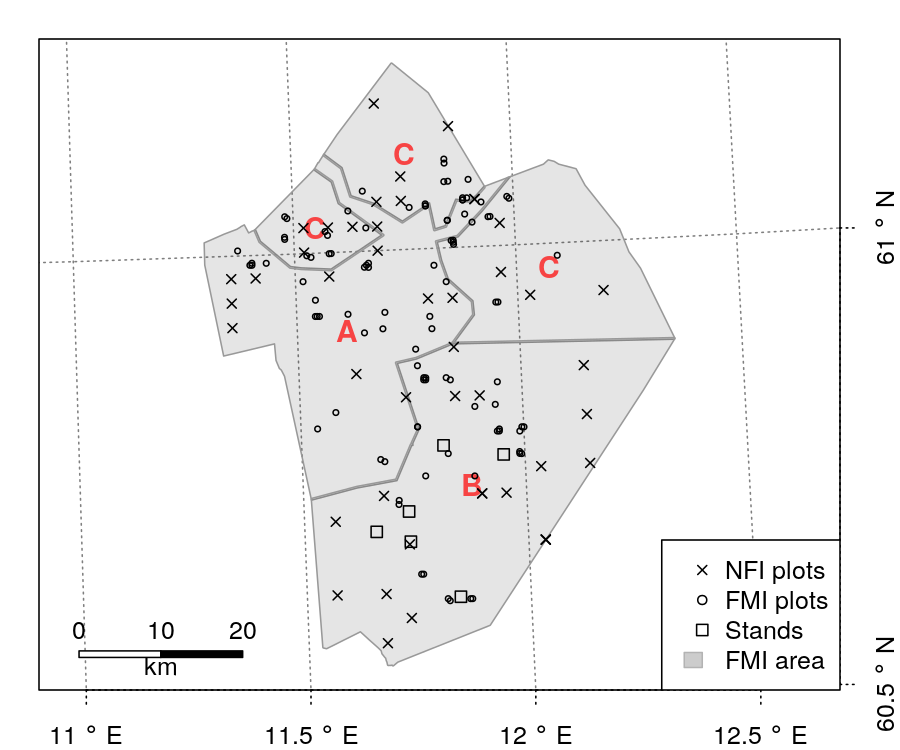}
\caption{Approximate locations of the NFI and FMI plots within the FMI area
and stands with a detailed inventory. Subdivisions
indicate different ALS projects (A-C). \label{fig_map2}}
\end{figure}

\subsection{Airborne laser scanning
data}\label{airborne-laser-scanning-data}

The ALS data used in this study were acquired as part of a national program to
create a fine-resolution and nationwide elevation model \citep{Kartverket2018}.
An available FMI \citep[unpublished]{mjosenskog2018} had been conducted within three ALS acquisition projects (Figure~\ref{fig_map2}, Table~\ref{tab_ALSacq}).
Additionally, data from other ALS acquisitions
in southeastern Norway (Figure~\ref{fig_map1}) with varying sensors, point
densities, and acquisition dates were used in the NFI models.

\begin{table}
\caption{Airborne laser scanning (ALS) data acquisitions in the study area.}
\label{tab_ALSacq}
\centerline{
\begin{tabular}[]{crrrr}
\toprule
ALS project & Municipality & Sensor & Point density (\(m^{-2}\)) &
Acquisition time\\
\midrule
A & Elverum & Optech Titan & 5 & May--June 2016\\
B & Våler & ALS70-HP & 2 & June--October 2016\\
C & Elverum & ALS70-CM & 2 & May--October 2016\\
\bottomrule
\end{tabular}
}
\end{table}

For each ALS project, ALS returns were extracted
that intersected with the NFI and FMI sample plots. We used the existing
national terrain model based on ALS data with a resolution of 1 m to
subtract terrain elevation from the return elevation to obtain heights
above ground, using bi-directional linear interpolation.
ALS height metrics were calculated from the point clouds: mean height (zmean),
height standard deviation (zsd), height percentiles (zp05, zp10,
zp20, \ldots{}, zp90, zp95).
Density metrics were calculated by dividing the distance between the lowest and the highest return into 10
bins with equal heights and calculating the percentage of returns above the lower bin threshold (d2, d3, ..., d10).
We calculated these metrics for first returns (*\_f) and for last returns (*\_l) without a height threshold, and with a height threshold of 2 m (*\_2m\_f, *\_2m\_l). 
For example, the 50th height percentile of the last returns with a height threshold of 2 meters would be abbreviated zp50\_2m\_l.
The percentage of all returns above 2 m (perc\_n\_2m) served as an additional density metric. 

For mapping purposes, we used 16 x 16 m grid cells, that correspond
approximately to the NFI and FMI plot sizes (see Sections~\ref{section_FMI} and~\ref{nfi-data}). For each grid cell we
extracted the ALS returns and subtracted the terrain height.
Subsequently, the same ALS metrics as for the sample plots were
calculated wall-to-wall for
each ALS project.

\subsection{FMI data}\label{section_FMI}

Forest stand polygons were available as part of the FMI \citep{mjosenskog2018}; they were delineated by visual
interpretation of remotely sensed data and classified into five strata based on tree
species, site index and maturity class. A total of 20,427 stands with
areas between 0.03 and 25 ha, covering a total area of 19,800 ha, were
within the mature spruce forest stratum. Within these
polygons, 402 systematically distributed sample plots in clusters with $3 \times 3$ plots were measured if they were located in spruce or pine forest. A total of
101 of these sample plots were located in mature spruce stands (Figure~\ref{fig_map2}). The field campaign took place in 2017.

The FMI sample plots are circular and have an area of 250 $m^{2}$. All
trees with a diameter at breast height (dbh) \(\geq\)~10~cm were registered. Heights were measured
for approximately 10 trees per plot. 
From these tree measurements, we estimated plot-level timber volume following the methodology of the
Norwegian NFI based on species specific volume models \citep{braastad1966volume,Brantseg1967,Vestjordet1967}.
For a detailed description of the volume estimation from tree-level measurements see \citet[Section 5.2.1.1]{breidenbach2020century}.
By convention, we speak of \textit{plot-level measurements} although we are aware of the model-related uncertainty.

The mean timber volume of the FMI was 51\% and 13\% larger than the mean timber volume of the NFI plots in the
FMI area and the whole study area, respectively (Table~\ref{tab_inventory_descr}).

\begin{table}
\caption{Summary of the plot-level timber volume measurements. SD is standard deviation and CV is coefficient of variation.}
\label{tab_inventory_descr}
\centerline{
\begin{tabular}[]{lrrrrrrr}
\toprule
Inventory & Year & n plots & \multicolumn{5}{c}{Timber volume (\(m^3 ha^{-1}\))}\\
\cmidrule(lr){4-8}
& & & Minimum & Mean &  SD & CV (\%)  & Maximum\\
\midrule
FMI & 2017 & 98 & 71 & 321  & 182 & 57 & 972\\
NFI in FMI area & 2014-2018 & 40 & 37 & 213 & 138 & 65 & 641\\
NFI & 2014-2018 & 244 & 29 & 285 & 167 & 59 & 884\\
\bottomrule
\end{tabular}
}
\end{table}

\subsection{FMI models}\label{fmi-models}

\textit{FMI models} were fitted following the methodology of forest management planning in Norway
\citep{naesset2014area} which is based on the area based approach \citep{Naesset2002}.
Log-log models were fitted for each ALS project using the estimated FMI plot timber volume 
measurements (Section~\ref{section_FMI}) as response and ALS metrics as explanatory
variables.

\begin{equation}
\ln vol_{j}=\beta_{0}+ \beta_{1}\ \ln\ x_1+ ... + \beta_{k}\ \ln\ x_k+\varepsilon_{j}, 
\label{eq_loglog}
\end{equation}
where \(vol_{j}\) is the timber volume at FMI plot \(j\) and $x_1,
\ldots{}, x_k$ are the explanatory variables. To correct for
transformation bias, half the model variance was added to the intercept
before transforming predictions back to the original scale.

Explanatory variables for the log-log models were selected using a
step wise forward and backward selection scheme informed by the Akaike
information criterion (AIC) (Table~\ref{tab_FMI_model_descr}). The
algorithm was allowed a maximum of four explanatory variables in
addition to the intercept to avoid overfitting.

Accuracies of the models were assessed based on leave-one-out cross-validation of the modeling data using root-mean-squared deviance (RMSD, RMSD\%) and mean deviance (MD, MD\%).

In the cases where models were evaluated using measured data RMSD and MD are the same as root-mean-squared error (RMSE) and mean error (ME). 
Here,  "R\(^2\) of Prediction" \citep[p.~152]{montgomery2012introduction}, further referred to as R\(^2\), is also given, allowing assessment of the variance in the predictions in relation to the variance in the population.
RMSD, MD, and R\(^2\) were defined as

\begin{equation}
{R\! M\! S\! D}=\sqrt{\frac{1}{n}\sum_{j=1}^{n}(y_{j}-\hat{y}_{j})^{2}},\qquad {R\! M\! S\! D\! \%}= \frac{R\! M\! S\! D\!}{\bar{y}}\cdot 100,
\label{eq_rmse}
\end{equation}

\begin{equation}
{M\! D}=\frac{1}{n}\sum_{j=1}^{n}(y_{j}-\hat{y}_{j}),\qquad {M\! D \%}=\frac{M\! D\!}{\bar{y}}\cdot 100,
\label{eq_md}
\end{equation}

\begin{equation}
{R^2}= 1 - \frac{\sum_{j=1}^{n}(y_{j}-\hat{y}_{j})^2}{\sum_{j=1}^{n}(y_{j}-\bar{y})^2},
\label{eq_r2}
\end{equation}

\noindent where \(y_{j}\) is the timber volume observed at plot \(j\),
\(\hat{y}_{j}\) is the back-transformed predicted timber volume at plot \(j\),
\(\bar{y}_{j}\) is the mean observed timber volume of all plots, \(n\)
is the total number of plots.

\begin{figure}
\centering
\includegraphics[width=\textwidth]{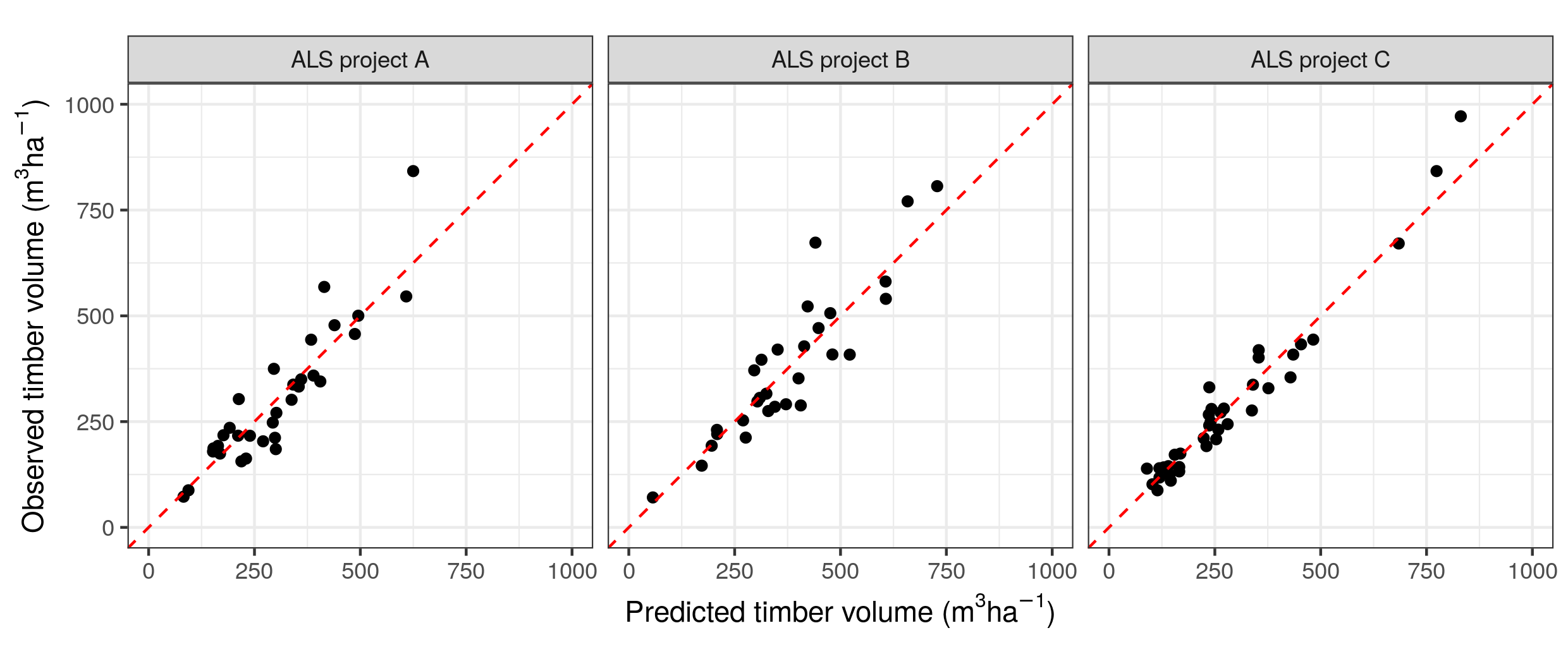}
\caption{Observed vs.~predicted timber volume using FMI models.
Predictions for the FMI plots were based on leave-one-out
cross-validation. \label{fig_CV_FMI_models_plots}}
\end{figure}

Across all ALS projects, the (cross-validated) RMSD was 62 \(m^3 ha^{-1}\) (19\%), MD 2 \(m^3 ha^{-1}\) (1\%) and R\(^2\) 0.88.
Figure~\ref{fig_CV_FMI_models_plots} illustrates the model fits for all
three ALS projects at plot level. RMSDs, MDs, and R\(^2\) using NFI data are given
in the Results (Section~\ref{model-comparison-at-plot-level}).

\begin{table}
\caption{Model parameter estimates, root-mean-squared deviance (RMSD) and mean deviance (MD) (based on leave-one-out cross-validation) of the FMI models. ALS is airborne laser scanning. ALS metrics are explained in detail in Section \ref{airborne-laser-scanning-data}.}
\label{tab_FMI_model_descr}
\centerline{
\begin{tabular}[]{lrlrrrrrrrr}
\toprule
ALS     &&                       &             &             &             &             &             & RMSD & MD & \\
project &n& Explanatory variables$^{\dagger}$ & \(\beta_0\) & \(\beta_1\) & \(\beta_2\) & \(\beta_3\) & \(\beta_4\) & (\%) & (\%) & R\(^2\)\\
\midrule
A & 32& Intercept, zmean\_f, d5\_2m\_f, d5\_2m\_l, zq20\_l   & 0.65 & 2.01 & -1.35 &  0.96 & -0.05 & 15 & 0 & 0.88\\
B &29& Intercept, zmean\_l, zq10\_f, zq10\_l                 & 2.26 & 1.70 & -0.10 &  0.15 &       & 22 & 1 & 0.90\\
C &37& Intercept, zmean\_f, zmean\_l, d9\_2m\_f, d9\_l       & 2.62 & 1.01 &  0.55 &  0.42 & -0.31 & 20 & 1 & 0.94\\
\bottomrule
\multicolumn{10}{l}{
\begin{minipage}[l]{\textwidth}
$^{\dagger}$ zmean = mean return height; 
zq* = *th percentile of return heights; 
d* = proportion of returns above the *th height bin; 
*\_f =  only using first returns; *\_l =  only using last returns; 
*\_2m = applying a minimum return height threshold of 2~m
\end{minipage}

}
\end{tabular}
}
\end{table}

\subsection{NFI data}\label{nfi-data}

The 815 permanent sample plots of the Norwegian NFI \citep{breidenbach2020century} in the study area were measured between 2014 and 2018 (Figure~\ref{fig_map1}). A total of 244 of the plots belonged to the mature spruce stratum (see Table~\ref{tab_inventory_descr}) of which 12 were in ALS project A, 17 in B, and 13 in C (Figures~\ref{fig_map1} and~\ref{fig_map2}). 
The NFI sample plots are circular with an area of 250~$m^2$.
On the sample plots, dbh and tree species of all trees
with dbh \(\geq\) 5 cm are registered. Tree heights are measured from
approximately 10 trees per plot, which are sampled using weights based
on diameter and distance to the plot center.
Positions of the sample plot centers were measured using differential GPS and GLONASS.

Like for the FMI data, timber volume was estimated following the methodology of the
Norwegian NFI (see Section~\ref{section_FMI}).
The timber volumes were fore- and back-casted to summer 2017, when the FMI was conducted.
To do so, we calculated the yearly increment from the recent two tree-level volume predictions
and multiplied the increment by the time difference which was then added
to the predicted timber volume at tree level. To adjust the NFI
data to the FMI protocol, trees with dbh \textless{} 10 cm were
discarded and the plots were stratified according to the FMI
stratification. Single tree volumes were subsequently totaled at plot
level, and estimates of timber volume per ha were calculated.

\subsection{NFI model}\label{nfi-model}

The bases for the timber volume map within the NFI-based Norwegian forest resources map SR16 \citep{Astrup2019} are linear mixed-effects models fitted to data from NFI sample plots of a larger region covered by available ALS data.
Tree species specific models are fitted with timber volume as the response and ALS metrics as fixed effects.
The ALS metrics used in the models are mean first return height above ground and a density metric.
An identifier at ALS-project level is used as random effect to address differences in the sensors and the acquisition conditions.
Because the within-group variance is observed to increase linearly with volume, heteroscedasticity is modeled using a variance function.

We re-fitted a linear-mixed effects model for timber volume of mature spruce stands using the ALS and NFI data within the study area with the adjusted dbh threshold of 10~cm.
We refer to this model as the \textit{"NFI model"}.
Fixed effects were zmean\_f and zmean\_f$^{2}$, as well as perc\_n\_2m as a density metric.
Before model fitting, 10 outliers that were likely affected by harvests were visually identified
and removed by analyzing residual plots.
A random effect on the ALS-project level was used for the slope of zmean\_f.

The NFI model was formulated as
\begin{equation}
\begin{split}
vol_{ij}=\beta_{0}+ ( b_{i}+\beta_{1} )\ zmean\_f_{ij}+\beta_{2}\ zmean\_f_{ij}^2+\beta_{3}\ perc\_n\_2m_{ij}+\varepsilon_{ij}, \\
i=1,...,m,\quad j=1,...,n_{i},\quad b_{i}\sim N(0,\sigma_{b}^{2}),\quad \varepsilon_{ij}\sim N(0,\sigma_{\varepsilon}^{2}{zmean\_f_{ij}}),
\end{split}
\label{eq_lme}
\end{equation}

\noindent where \(vol_{ij}\) is the timber volume at sample plot \(j\) in ALS
project \(i\). \(\beta_1,\beta_2,\beta_3\) are the fixed-effects
parameters, \(b_{i}\) is the random-effect parameter, \(n_i\) is the
number of sample plots within ALS project \(i\), \(m\) = 26 is the
number of ALS projects, \(\sigma_{b}^{2}\) is the variance of the random
effect, and \(\sigma_{\varepsilon}^{2}\) is the residual variance.
The variance function has a single variance covariate zmean\_f. 
The model was fitted using the nlme package \citep{nlme} in R
\citep{rlanguage}.

\begin{figure}

\includegraphics[scale=0.7]{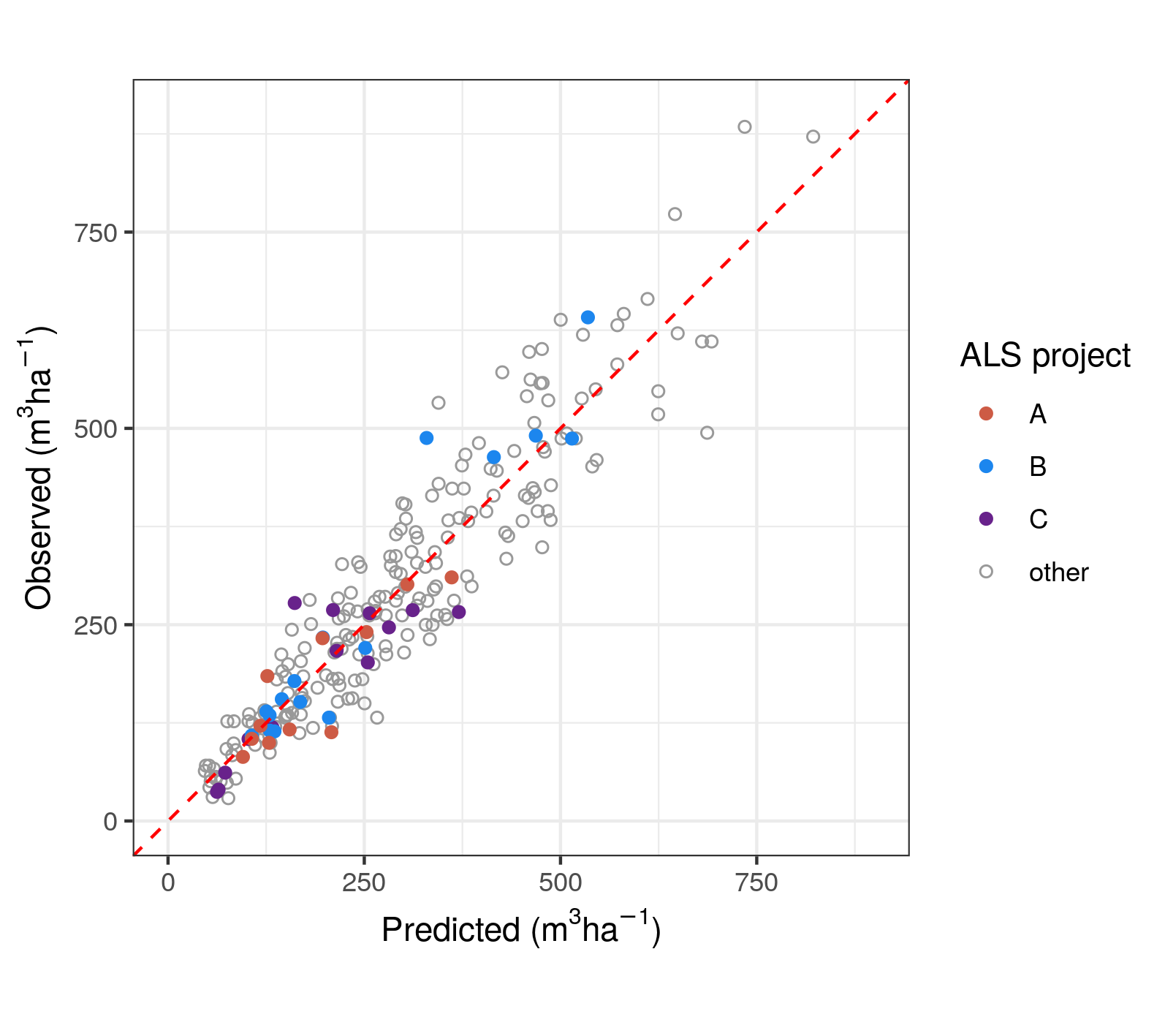}
\caption{Observed~vs.~predicted timber volume (\(m^3 ha^{-1}\)) at the
NFI plots using the NFI model. Predictions are based on
leave-one-out cross-validation. \label{fig_FMI_models_plots}}
\end{figure}

The estimated parameters and other characteristics of the NFI model are
shown in Table~\ref{tab_model_params}. The cross-validated RMSD of the NFI model was
21\% with no systematic deviation (MD) based on all NFI sample plots and the R\(^2\) was 0.88.
The RMSD on plot level describes the uncertainty of the NFI-based map on
pixel level. The observed vs.~cross-validation-predicted timber volume is shown in
Figure~\ref{fig_FMI_models_plots}. RMSDs, MDs, and  R\(^2\) using FMI data are
given in the Results (Section~\ref{model-comparison-at-plot-level}).

\begin{table}
\caption{Parameter estimates of the mixed-effects model.
\label{tab_model_params}}
\centerline{
\begin{tabular}[]{lrrrrrrr}
\toprule
& \(\beta_0\) & \(\beta_1\) & \(\beta_2\) & \(\beta_3\) & \(\sigma_{b}\)
& \(\sigma_{\varepsilon}\) & \(b_{i}\)\\
\midrule
Parameter estimates & 23.73 & 19.75 & 0.98 & -63.34 & 1.39 & 17.08 &
A:-0.92, B:1.45, C:-0.75\\
\bottomrule
\end{tabular}
}
\end{table}

\subsection{Adjusted NFI model: utilizing additional local sample plots
for model improvement}
\label{improved-nfi-model-utilizing-additional-local-sample-plots-for-model-improvement}

We attempted to improve the local accuracy of the NFI based map by extending the modeling data set of the NFI model with local sample plots.
Two data sets were tested: 
(i) a combination of the NFI and all FMI sample plots and 
(ii) a combination of the NFI and a subset of FMI sample plot data.
The NFI models were refitted with the extended modeling data using Equation~\ref{eq_lme}.
The subset of FMI sample plots was chosen based on the value of zmean\_f. 
A comparison of the NFI and FMI sample plots shows the presence of relatively more plots with larger volumes in the FMI data (Table~\ref{tab_inventory_descr}).
Because of the correlation of zmean\_f and timber volume, the FMI sample plots were sorted based on the value of zmean\_f, and a number of plots were chosen from the top of this list, i.e.~plots with the largest values of zmean\_f.
By iterative testing and evaluating the improvement, the number of FMI sample plots selected to be combined with the NFI data was set to 7 per ALS project.
Predictions of timber volume at the FMI sample plots for the calculation of accuracy measures were obtained using leave-one-out cross-validation for sample plots that were used in the modeling data.
In the following, we refer to the resulting models as \emph{"adjusted NFI models"}.

\subsection{Comparison of the FMI and NFI models using stand-level estimates}
\label{comparison-of-the-fmi-and-nfi-models-using-stand-level-estimates}

Using the fitted models, timber volume was predicted for all 16 x 16 m cells of a grid covering the study area. To obtain synthetic estimates of stand-level timber volume we averaged the predictions of the grid cells with a center within the stands.
To reduce the computational cost, we randomly selected 200 FMI stands per ALS project. 
Stands were required to have a minimum area of 1 ha.
An area-to-perimeter ratio can describe the compactness of stands.
Only stands with a ratio \(\sqrt{area}/perimeter\) \textgreater{}  0.2. were selected to reduce the edge effect of stand borders on the resulting estimates.
We compared the synthetic stand-level estimates of the FMI models with those of the NFI models using RMSD, MD, and R\(^2\) with \(y_j\) and \(\hat{y}_j\) as FMI and NFI model synthetic estimate, respectively.

\subsection{Comparison with an independent forest inventory}
\label{section_stand_inventory}

As basis for a comparison with independent data, 61 sample plots in six forest stands were measured in ALS project B in 2018.
The stands were selected from the delineated forest stand polygons with areas between 1.5 and 5.5~ha and a ratio \(\sqrt{area}/perimeter\)~\textgreater{}~0.2 (see Section~\ref{comparison-of-the-fmi-and-nfi-models-using-stand-level-estimates}).
10 to 11 sample plots were randomly selected from nodes of a $20 \times 20$~m grid intersecting with the stand boundaries.
After the removal of one plot that coincided with forest road, one stand had nine sample plots, resulting in a total of 60 sample plots for the comparison.

In the field, plot center positions were measured using handheld GPS devices.
The plots were circular with an area of 250~$m^2$ and were measured according to the NFI protocol (Section~\ref{nfi-data}).
The tree measurements were adjusted to the FMI protocol by discarding trees with dbh~$<$~10~cm, and single tree volumes were estimated following the methodology of the Norwegian NFI \citep{breidenbach2020century}.
Individual tree volumes were totaled at plot level, and per-ha volume was calculated.
Subsequently, stand-level estimates were obtained by averaging plot-level measurements \citep[see][]{rahlf2014comparison}.
Stand-level estimates based on these measurements are referred to as \textit{direct estimates}.
Direct estimates were treated as an observation.

Plot-level measurements ranged from 84 to 646~\(m^3 ha^{-1}\) and stand-level direct estimates from 154 to 380~\(m^3 ha^{-1}\).
ALS data were clipped to the plot boundaries and plot-level timber volume was predicted using the area based approach (see Section~\ref{fmi-models}) for each plot.
RMSD, MD, and R\(^2\) were calculated using Equations~\ref{eq_md}--\ref{eq_r2} by substituting \(y_j\) with the direct estimate and \(\hat{y}_j\) with the average prediction for the plots within a stand using the FMI, NFI, or adjusted NFI model. It may be noted that there is no sampling-related uncertainty in the direct estimate due to this procedure.

\section{Results}\label{results}

\subsection{Comparison of FMI and NFI predictions and estimates\label{model-comparison-at-plot-level}}

\begin{figure}
\centering
\includegraphics[scale=0.8]{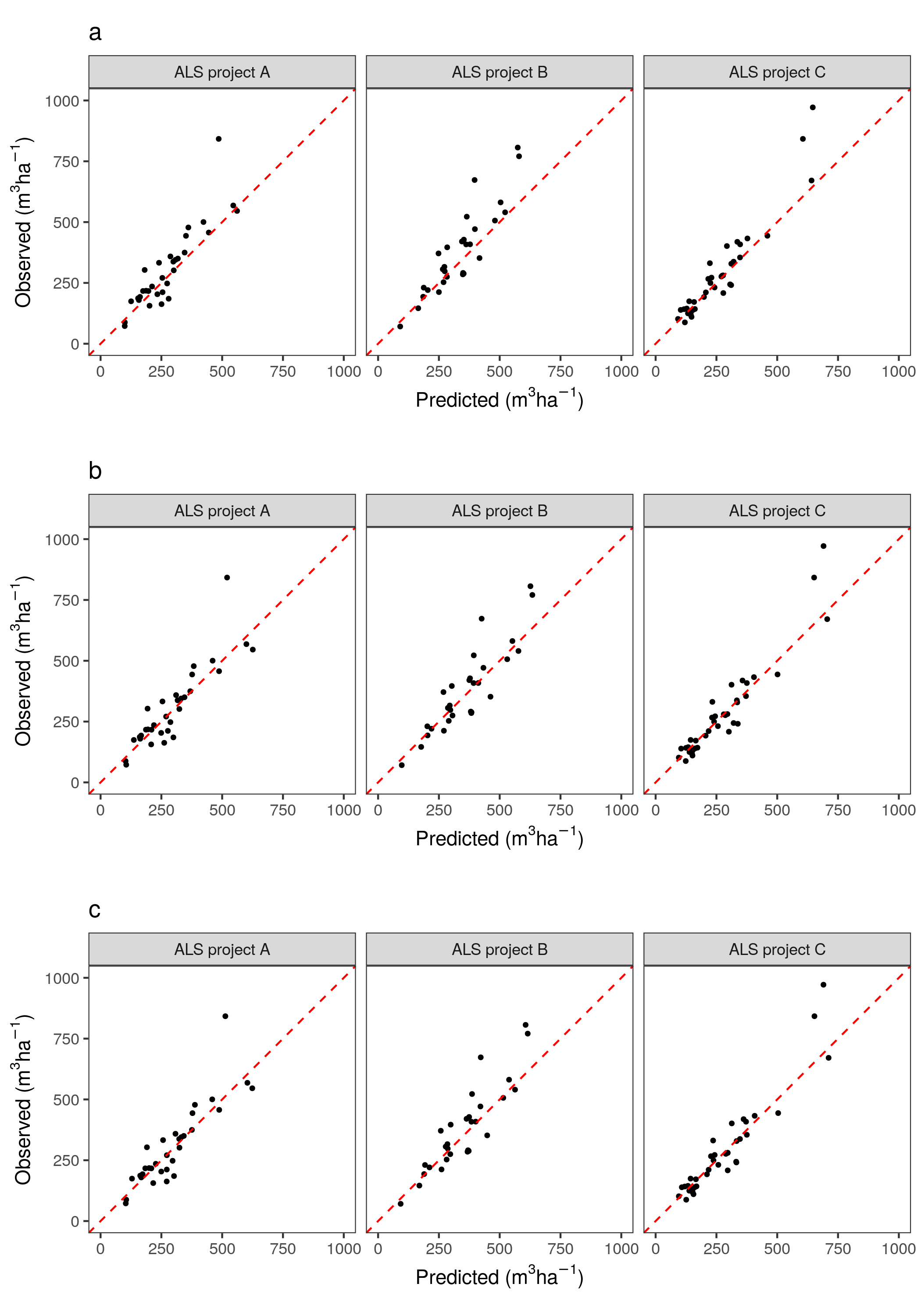}
\caption{Observed vs.~predicted timber volume (\(m^3 ha^{-1}\)) at the
FMI sample plots; predictions derived from the NFI model based on (a)
NFI data, (b) NFI and all FMI data, and (c) NFI data and the 7 FMI
sample plots with the largest zmean\_f. \label{fig_plot_predictions}}
\end{figure}

We validated the FMI models using NFI plots in the FMI area resulting in a RMSD and MD of 26\% (56 \(m^3 ha^{-1}\)) and -4\% (-8 \(m^3 ha^{-1}\)), respectively (R\(^2\)~=~0.83).
Similarly, the NFI model was validated using FMI plots, resulting in an RMSD and MD of 27\% (86~\(m^3 ha^{-1}\)) and 10\% (33~\(m^3 ha^{-1}\)), respectively (R\(^2\)~=~0.78) (Table~\ref{tab_added_FMI_plots_eval_meas}).
The NFI model showed a slight tendency to underpredict timber volume for FMI plots with timber volume \textgreater{} 350 \(m^3 ha^{-1}\), which was mostly visible in ALS project B but less so in ALS projects A and C (Figure~\ref{fig_plot_predictions} a).

To analyze if the NFI-based model could be improved by using additional local sample plots, we fitted models based on a combination of the NFI data and either all or a subset of the FMI data (see Section~\ref{improved-nfi-model-utilizing-additional-local-sample-plots-for-model-improvement}).
Models fit to either of the combinations of NFI and FMI data resulted in decreased RMSDs and MDs in all ALS projects compared to the NFI model (Table~\ref{tab_added_FMI_plots_eval_meas}, Figure~\ref{fig_plot_predictions}), when using FMI data for validation.
The largest improvements of RMSDs were achieved when adding all FMI sample plots to the modeling data.
The RMSD decreased by 3 and MD by 6 percentage points, nearly reducing the systematic error to a third of the MD of the NFI model.
The RMSD of the adjusted NFI model, where only a subset of the FMI sample plots was added to the NFI data, had similar properties as the model using all NFI and FMI plots, with even smaller MDs in ALS projects A and C.

\begin{table}
\caption{Differences between NFI model predictions and measurements at
the FMI sample plots. Values are based on independent validation
for NFI data and cross-validation for FMI data. ALS is airborne laser scanning, RMSD is root-mean-squared deviance, and MD is mean deviance.}
\label{tab_added_FMI_plots_eval_meas}
\begin{tabular}[]{llrrrrr}
\toprule
Modeling data & ALS project & RMSD (\(m^3 ha^{-1}\)) & RMSD\% & MD
(\(m^3 ha^{-1}\)) & MD\% & R\(^2\)\\
\midrule
NFI & ALS project A & 83.14 & 27 & 30.79 & 10 & 0.73\\
NFI & ALS project B & 96.56 & 25 & 46.08 & 12 & 0.69\\
NFI & ALS project C & 78.38 & 27 & 25.24 & 9 & 0.83\\
NFI & total & 85.65 & 27 & 33.22 & 10 & 0.78\\
\addlinespace
NFI \& FMI & ALS project A & 76.69 & 25 & 11.25 & 4 & 0.77\\
NFI \& FMI & ALS project B & 84.11 & 22 & 16.60 & 4 & 0.76\\
NFI \& FMI & ALS project C & 69.56 & 24 & 8.46 & 3 & 0.87\\
NFI \& FMI & total & 76.43 & 24 & 11.78 & 4 & 0.82\\
\addlinespace
NFI \& top 7 FMI & ALS project A & 78.25 & 26 & 9.61 & 3 & 0.76\\
NFI \& top 7 FMI & ALS project B & 86.55 & 23 & 27.28 & 7 & 0.75\\
NFI \& top 7 FMI & ALS project C & 69.70 & 24 & 7.11 & 2 & 0.87\\
NFI \& top 7 FMI & total & 77.78 & 24 & 13.90 & 4 & 0.81\\
\bottomrule
\end{tabular}
\end{table}

We compared synthetic estimates of the FMI models, the NFI model, and
the improved NFI models for a sub-sample of forest stands in the FMI area. As for the predictions at plot level, estimates of the NFI model
tended to be smaller than estimates of the local FMI models. For stands with
small timber volume the NFI model produced generally larger estimates.
On average, this resulted in a MD of 8\%. Estimates deviated least in
ALS project C and most in A, regardless of the combination of sample
plot data used for fitting the NFI model (Table
\ref{tab_splan_predictions}). 

\begin{figure}
\centering
\includegraphics[scale=0.8]{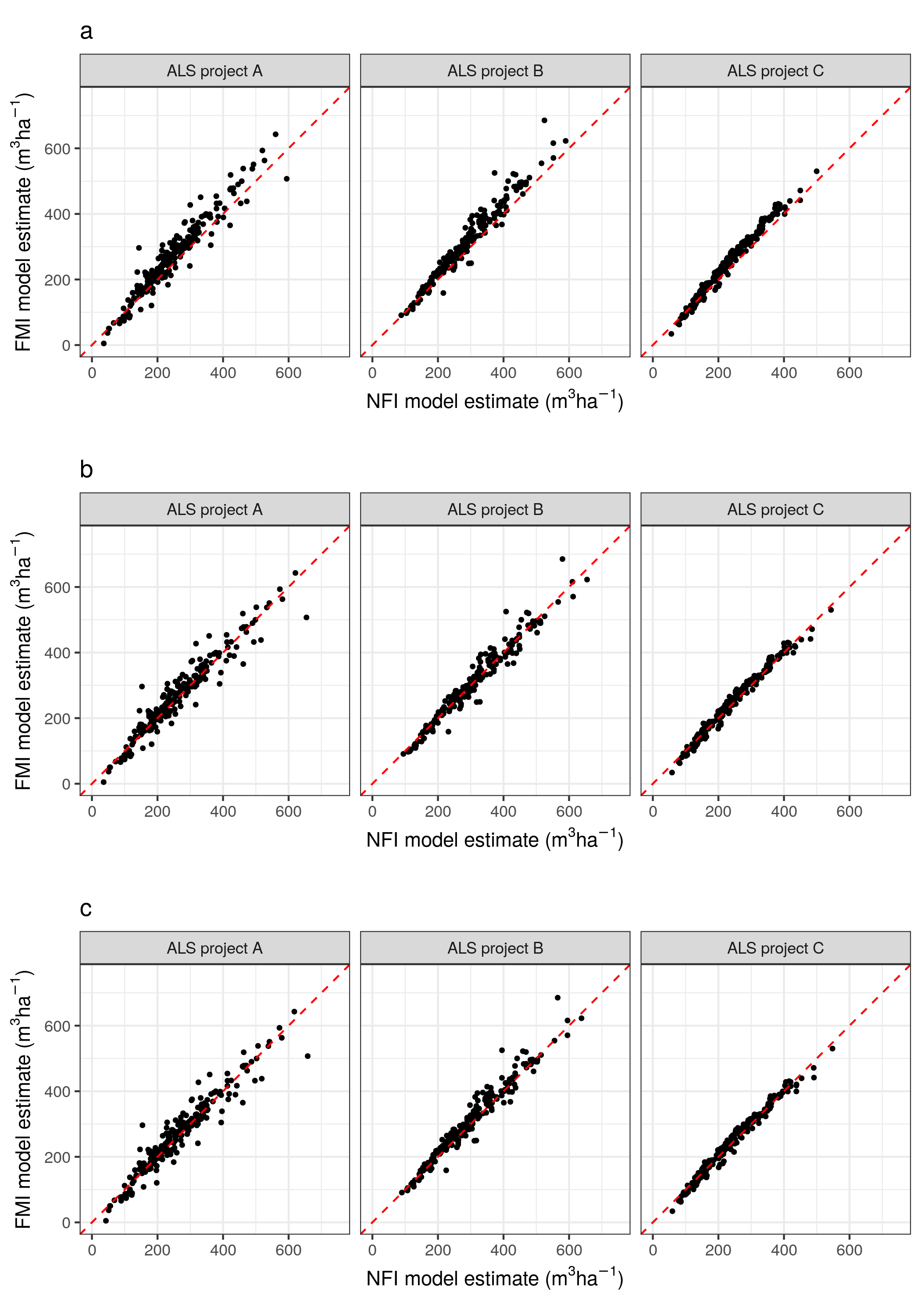}
\caption{Comparison of synthetic stand-level timber volume estimates
(\(m^3 ha^{-1}\)) resulting from the FMI and NFI models. NFI
model estimates are derived from models (ref) based on (a) NFI data, (b)
NFI and all FMI data, and (c) NFI data and the 7 FMI sample plots with
the largest zmean\_f. \label{fig_splan_predictions}}
\end{figure}

The addition of local sample plots to the NFI data for the model fit reduced the differences between estimates based on the NFI and FMI models, resulting in a smaller RMSD and especially smaller MD of the adjusted NFI models, compared to the original NFI model.
While the largest reduction was achieved when using all FMI sample plots, the reduction
for the adjusted NFI model with 7 additional FMI sample plots was
similar, with differences in RMSD and MD of 0--1 percentage points. The
smaller deviation of the estimates of the adjusted NFI models from the
estimates of the FMI models can be seen in Figure
\ref{fig_splan_predictions}.

\begin{table}
\caption{Differences between synthetic stand estimates of the FMI models and the NFI models based on 200 randomly selected stands. ALS is airborne laser scanning, RMSD is root-mean-squared deviance, and MD is mean deviance.} \label{tab_splan_predictions}
\begin{tabular}[]{llrrrr}
\toprule
Modeling data & ALS project & RMSD (\(m^3 ha^{-1}\)) & RMSD\% & MD
(\(m^3 ha^{-1}\)) & MD\%\\
\midrule
NFI & A & 40.25 & 15 & 23.67 & 9\\
NFI & B & 37.55 & 12 & 25.95 & 8\\
NFI & C & 22.83 & 10 & 15.72 & 7\\
NFI & All & 34.60 & 13 & 21.86 & 8\\
\addlinespace
NFI \& FMI & A & 32.56 & 12 & 6.35 & 2\\
NFI \& FMI & B & 24.75 & 8 & 1.68 & 1\\
NFI \& FMI & C & 13.75 & 6 & 2.90 & 1\\
NFI \& FMI & All & 25.16 & 9 & 3.71 & 1\\
\addlinespace
NFI \& top 7 FMI & A & 32.91 & 12 & 4.22 & 2\\
NFI \& top 7 FMI & B & 27.15 & 9 & 10.85 & 4\\
NFI \& top 7 FMI & C & 13.91 & 6 & 0.57 & 0\\
NFI \& top 7 FMI & All & 26.14 & 10 & 5.21 & 2\\
\bottomrule
\end{tabular}
\end{table}

\subsection{Comparison with independent data}\label{comparison-standinv}

\begin{figure}
\centering
\includegraphics[width=\textwidth]{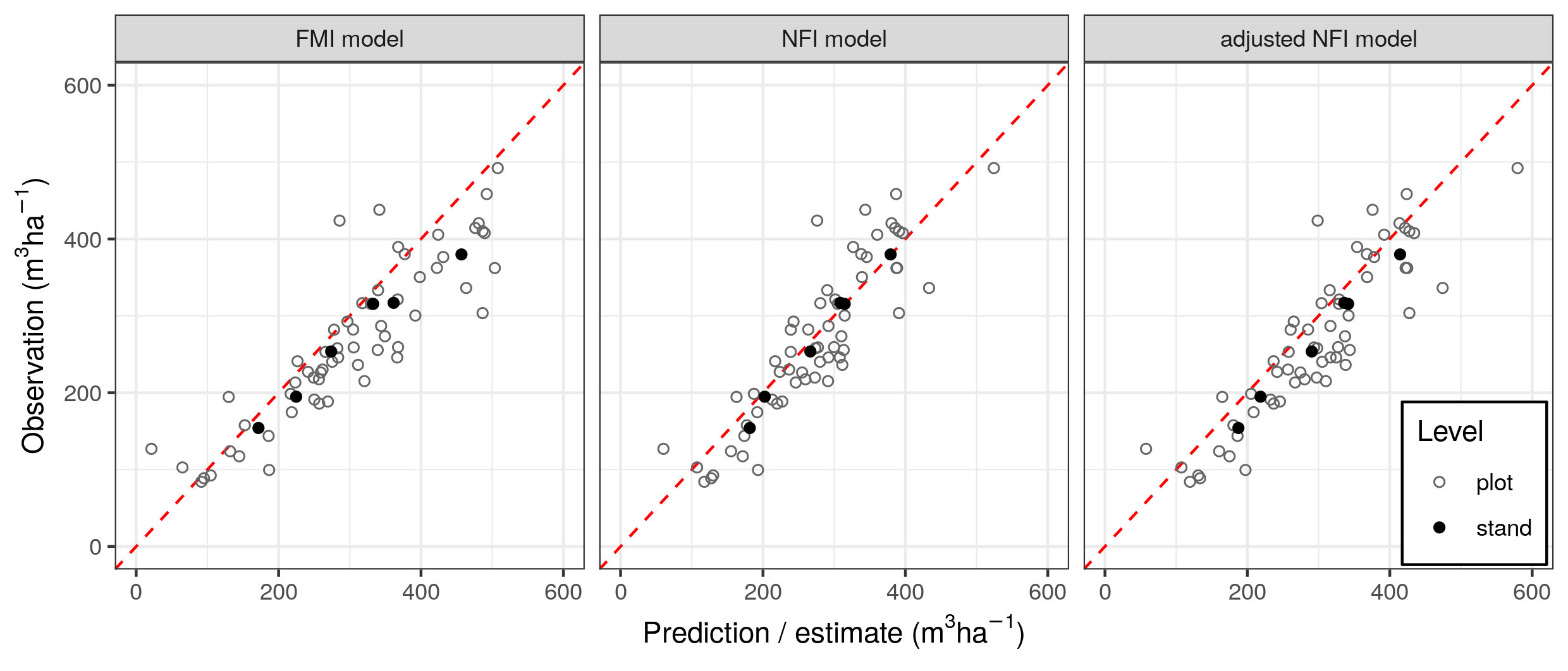}
\caption{Comparison of plot-level predictions and stand-level estimates to an independent forest inventory in ALS project B. \label{fig_validation_inventory}}
\end{figure}

The FMI model, the NFI model and the adjusted NFI model predictions were validated using an independent forest inventory in ALS project B.
Because the adjusted NFI model fitted with a subset of FMI plots was shown to be similarly accurate as the adjusted NFI model fitted with all NFI plots (Section~\ref{model-comparison-at-plot-level}), only the former was used in this analysis.
Of the three models, the NFI model performed best at stand and at plot level (Figure~\ref{fig_validation_inventory}).
At plot level, the NFI model and the adjusted NFI model produced smaller RMSDs than the FMI model (Table~\ref{tab_validation_inventory}).
Similar results were obtained at stand level, when the sample plot measurements and predictions were averaged for all plots within each of the 6 stands; RMSDs and MDs were smallest for the NFI and adjusted NFI model prediction.
RMSDs were smaller at stand level than at plot level, while MDs had similar values.

\begin{table}
\caption{Differences between plot and stand-level estimates of the NFI and FMI models and direct estimates of an independent inventory. RMSD is root-mean-squared deviance and MD is mean deviance.} \label{tab_validation_inventory}

\begin{tabular}{llrrrrrr}
\toprule
Level & Model & n & RMSD (\(m^3 ha^{-1}\)) & RMSD\% & MD (\(m^3 ha^{-1}\)) & MD\% & R\(^2\)\\
\midrule
plot & FMI model & 60 & 65 & 24 & -34 & -13 & 0.59\\
plot & NFI model & 60 & 48 & 18 & -7 & -2 & 0.78\\
plot & adjusted NFI model & 60 & 55 & 21 & -29 & -11 & 0.71\\
\addlinespace
stand & FMI model & 6 & 40 & 15 & -34 & -13 & 0.72\\
stand & NFI model & 6 & 13 & 5 & -6 & -2 & 0.97\\
stand & adjusted NFI model & 6 & 30 & 11 & -29 & -11 & 0.85\\
\bottomrule
\end{tabular}
\end{table}

\section{Discussion}\label{discussion}

We compared the accuracies of a NFI model, as it is used in a NFI-based map linking ALS data and NFI sample plots, and local FMI models for timber volume of mature spruce forest.
The accuracy of the NFI model was similar to that of the FMI models at plot and stand level.
The NFI model was slightly more accurate than the FMI models, when an independent dataset was used for validation.
Here, the NFI model also showed a smaller systematic error and was able to explain a larger part of the variation than the FMI and the adjusted NFI models.

In a direct comparison of stand level estimates, differences between the FMI and NFI models could be observed but it is unknown whether the FMI or NFI models are closer to reality.
However, the differences between the FMI and NFI models were relatively small and had the same magnitudes as typical stand-level RMSDs of ALS models \citep{Naesset2004d}.
While no clear improvement of the NFI model accuracy could be observed when using additional local sample plots, it reduced the deviations of stand-level estimates based on the NFI model and the FMI models, making the model estimates of the NFI model more similar to the estimates of the FMI model.

MDs of the NFI model at plot level, as well as the RMSD of the estimates
of the stands with validation inventory, were in line with previously
reported RMSDs for ALS data \citep{Naesset2004d, rahlf2014comparison}.
Other studies on NFI-based maps  for timber volume using ALS data reported similar or lower plot-level accuracies \citep{nord2012estimation, monnet2016wide, nilsson2016nationwide}.
Analyzing the use of NFI plots in ALS based forest management inventories, \citet{Maltamo2009a} reported a larger stand-level RMSD of 19.66\% and a similar MD of -1.13\%, compared to the NFI model accuracies.

Testing the application of a regional model to smaller areas, \citet{noordermeer2019comparing} fitted local and regional multiplicative regression models using data from multiple FMIs.
Accuracies for timber volume in productive mature forests were slightly higher for local than for regional models with RMSDs similar to our study.
\citet{tompalski2019demonstrating} analyzed the
transferability of ALS models by predicting forest parameters using
different ALS data within the same study area in Canada.
Unlike our findings, they reported only minor changes in RMSD and MD for ordinary
least square models on plot level using cross-validation.
However, differences in the ALS data were simulated by reducing point densities, which has been reported to have only limited effects on model accuracies \citep{Gobakken2007, jakubowski2013tradeoff}.

We used two approaches to compare accuracies of the models: Using the training data of the other model as validation data with cross-validation for the additional sample plots of the adjusted NFI models, and using an independent forest inventory as reference data set.
The independent forest inventory allowed a comparison of accuracies based on the same reference data.
Since the independent forest inventory was located in only one ALS project, the use of the FMI data served to assess differences in the NFI model accuracies between the ALS projects.
A reason for the differences between accuracies obtained with the two approaches might be the distribution of the response variable in the data.
While the measured volumes of the independent forest inventory vary around the center of the distribution of timber volume of the NFI data, the FMI sample plots show a larger range, extending over the maximum timber volume of the NFI sample plots.
The largest difference between the model validation using cross-validation and an independent forest inventory was observed in the MD value.
While these differences could be observed, both validation approaches revealed in general the same tendencies.

The validation of the NFI-based map in this study is based on the comparison with a single FMI across three ALS projects.
As the differences in accuracies of the NFI model between the ALS projects suggest, an exhaustive analysis of differences between NFI and FMI accuracies would need to include multiple FMIs from different regions.
Similarly, the interpretation of stand-level accuracies should therefore be carried out carefully.
Stand-level accuracies in this study are based on 60 plots within six forest stands.
However, even though estimation of stand-level parameters is the aim of remote sensing applications in FMIs, the analysis of plot-level accuracy is common in FMI research studies \citep{naesset2014area}.
In addition, the increase of accuracy when aggregating plot-level predictions at stand-level is in line with previous studies \citep{rahlf2014comparison, bohlin2017mapping}.

To analyze the influence of a few additional FMI plots on the NFI model, plots were selected based on their value of zmean\_f.
The result showed that these plots slightly improved the NFI model accuracy and reduced the systematic error only for plot-level predictions when FMI data were used for validation, and the improvement differed between the ALS projects.
A reason might be non-optimal selection of the local plots.
Methods based on sample distributions and local conditions \citep{grafstrom2012spatially} could be considered to improve the sample plot selection.

In this study, time differences in the data were relatively small.
However, in the case of large time differences between the acquisition of remotely sensed data and the field campaign for local model adjustment, map predictions would need to be updated to avoid the introduction of bias.
Besides newly available remotely sensed data, growth modeling could be used for updating.
The shelf life of ALS data is expected to be longer than 10 years \citep{mcroberts2018shelf}, longer than traditional inventory cycles of FMIs \citep{maltamo2020comprehensive}.
While we used solely Norwegian data in this case study, we assume the results to be comparable in countries where environmental conditions, forest management and inventory systems are similar.

This study provides a first analysis of the use of predictions from NFI-based maps in FMIs.
We concentrated on mature spruce stands, because this stratum includes the most valuable stands and is of the highest interest for forest managers.
Further work will need to explore the application of NFI-based maps in FMIs covering all development classes and also other tree species.
Additionally, more parameters, e.g. stem numbers or mean tree height, need to be tested for use in an operational FMI.

In conclusion, the presented comparison shows that estimates using a NFI-based map can be similar to or more accurate than estimates from a FMI.
The similar accuracies indicate that NFI-based maps can directly be used in FMIs for timber volume estimation in mature spruce stands, leading to potentially large cost savings with only small, if any, losses in accuracy.
The addition of local sample plots when fitting an NFI model did not clearly improve the model accuracy.

\section*{Acknowledgements}\label{acknowledgements}

The authors are grateful to the anonymous reviewers whose comments greatly helped to improve the clarity and precision of the paper.
We thank Glommen-Mjøsen Skog SA, for the provision of FMI data.
This study was supported by the Norwegian Forest Development Fund (Utviklingsfondet for skogbruket) under contracts \#16/66234-3 and \#19/42118-6, the Forest Initiative Fund (Skogtiltaksfondet) under contract \#B-2016-45, and the Norwegian Institute of Bioeconomy Research. 

\bibliographystyle{spbasic} 
\bibliography{bibliography.bib}

\end{document}